# Configuring Antenna System to Enhance the Downlink Performance of High Velocity Users in 5G MU-MIMO Networks

Md. Asif Ishrak Sarder
*Islamic University of Technology*
Gazipur-1704, Bangladesh.
Email: asifishrak@iut-dhaka.edu

Fehima Tajrian
*Islamic University of Technology*
Gazipur-1704, Bangladesh.
Email: Fehimatajrian@iut-dhaka.edu

Moontasir Rafique
*Islamic University of Technology*
Gazipur-1704, Bangladesh.
Email: moontasir@iut-dhaka.edu

Mariea Anzum
*Islamic University of Technology*
Gazipur-1704, Bangladesh.
Email: marieasharaf@iut-dhaka.edu

Abdullah Bin Shams
University of Toronto
Toronto, ON M5S 3G8, Canada.
Email: abdullahbinshams@gmail.com

*Abstract*— An exponential increase in the data rate demand, prompted several technical innovations. Multi User Multiple Input Multiple Output (MU-MIMO) is one of the most promising schemes. This has been evolved into Massive MIMO technology in 5G to further stretch the network throughput. Massive MIMO tackles the rising data rate with the increase in the number of antenna. This comes at the price of a higher energy consumption. Moreover, the high velocity users in MU-MIMO scheme, experiences a frequent unpredictable change in the channel condition that degrade it's downlink performance. Therefore, a proper number of antenna selection is of paramount importance. This issue has been addressed using machine learning techniques and Channel State Information (CSI) but only for static users. In this study, we propose to introduce antenna diversity in spatial multiplexing MU-MIMO transmission scheme by operating more number of reception antenna compare to the number of transmission antenna. The diversity improves the downlink performance of high velocity users. In general, our results can be interpreted for large scale antenna systems like Massive MIMO. The proposed method can be easily implemented in the existing network architectures with minimal complexity. Also, it has the potential for solving real-life problems like call drops and low data rate to be experienced by cellular users traveling through high-speed transportation systems like Dhaka MRT project.

## I. Introduction

Recent forecasts predict a massive leap of mobile data traffic in coming years, about seven-fold increment in between 2017 to 2022[1]. Such surge of data traffic calls for a better wireless communication technology. To meet such demand, Multi User Multiple Input Multiple Output (MU-MIMO) is one of the most promising schemes. Since first introduced in LTE Release 8 and further enhancements in the later releases[2], MU-MIMO has been offering a high data rate, low latency with better spectral efficiency by implementing features like beamforming, space-time coding and spatial multiplexing. All these has been incorporated to form Massive MIMO. In Massive MIMO, a multiantenna array is arranged at the base station, ensuring transmission of multiple independent data streams towards multiple users through various transmitting antennas. This is the spatial multiplexing transmission technique which improves the overall spectral efficiency[3].

Nevertheless, such robust framework of MU-MIMO often suffers from various small-scale hindrances like spatial interference and obstacles in setting up larger array of multiantenna system causing high power consumption, operating and CSI (Channel State Information) detection complexities. Some rigorous studies have already recognized certain issues and came-up with various schemes to ensure a better Quality of Service (QoS). A heuristic solution, proposed in [4], uses Particle Swarm Optimization (PSO) to specify minimum number of antenna elements for transmission purpose. Targeting sum-rate maximization, [5] has adopted SUS and JASUS algorithms with precoding technique for user selection and antenna scheduling, respectively. Another antenna selection method by analyzing CSI has been proposed in [6] to enhance energy efficiency (EE) of massive MIMO structure. Moreover, Deep Learning (DL) approach has been applied for optimum antenna selection to extract even better performance from the massive MIMO in [7].

To ensure ubiquitous connectivity of massive MIMO, all practical scenarios should be considered beforehand such as random user mobility especially ongoing projects like Mass Rapid Transit (MRT) system in various metropolitan cities including Dhaka city[8] where estimated velocity of the trains is expected to reach a maximum of 100km/h [9]. The Doppler shift induced from user mobility can cause sudden variation in channel condition and thus degrade the MIMO connectivity. However, the abovementioned research works did not consider user mobility in their proposed schemes. Also, DL approach for optimum antenna selection can cause higher latency during real-time implementations. Authors in [10-13] consider random user mobility during their investigation. For VANET technology, a joint user and antenna selection algorithm has been proposed in [10]. Whereas [11] has applied MMSE detector and RZF precoder for realistic performance analysis of MIMO. In our previous works [12-16] performance analysis between scheduling schemes under various network types and between different spatial multiplexing techniques have been observed under wide range of user mobility.

In this paper, we propose a simple approach of utilizing antenna diversity in spatial multiplexing technique by reducing the number of operating transmitter antennas than the receiver antennas to improve overall throughput for high-velocity users. To emulate a massive MIMO technology and circumvent the simulation complexities, a 4×4 MIMO system has been designed. The proposed method will enhance wireless network performance independent of the scheduler and transmission schemes for high velocity users. Our method can be easily implemented in the existing network architectures with minimal complexity. Also, it has the potential for solving real-life problems like call drops and low data rate to be experienced by cellular users traveling through high-speed transportation systems like Dhaka MRT project.

The rest of the paper is organized as follows. In Sec.2, various network parameters and network model are described briefly. Simulation model is introduced in Sec.3. Simulation Results are provided and discussed in Sec.4 and finally Sec.5 concludes the whole paper.

## II. SYSTEM MODEL

**1. Closed Loop Spatial Multiplexing (CLSM):**

In spatial multiplexing, multiple independent data streams are transmitted from multiple transmitting antennas to the individual users under a MIMO coverage area to improve spectral efficiency with higher data rate. For a given antenna configuration (X×Y), number of independent Data stream $M$, channel bandwidth $B$ and signal-to-noise ratio $S/N$, the equation for corresponding channel capacity $C$ is:

$$C = MB \log_2(1 + \frac{S}{N}) \quad (1)$$

Where the number of independent Data stream $M$ must be smaller than or equal to the minimum number of transmitting antenna ($X$) and receiving antenna ($Y$)[13].

In CLSM, after a successful data reception, a feedback is sent from UE to the base station incorporating parameters indicating channel condition. Here, UE interprets the received data stream and reports back CSI, RI and PMI. The CSI (Channel State Information) suggests the base station an optimum modulation scheme and code rate that should be used in downlink. The RI (Rank Indicator) implies the possible number of spatial layers a UE can support considering the channel quality. PMI (Precoding Matrix Index) is used to govern the base station to utilize one of the precoding matrices for DL (Downlink) based on the received CSI feedback. For the four transmit antenna configuration in CLSM, the LTE codebook comprises 16 precoders. For simulation, 16-QAM is used in this paper.

**2. Network Model:**

For the simulation purpose, to implement massive MIMO using spatial multiplexing and a variation in antenna diversity, a conventional wireless network using multiple macro cells has been considered as delineated in Figure 1. The model comprises a two-tier network with 19 macro cell BSs aligned in a hexagonal manner. The distance between two BSs ($I_{BS}$) is kept 500 m. Each macro cell consists tri-sector antennas, outlined using antenna model Kathrein 742215[18].

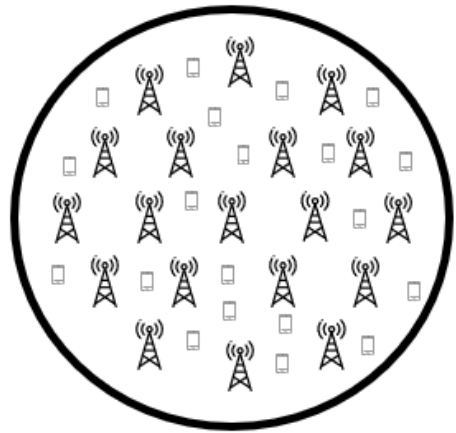

Fig.1: A two-tier wireless network using macro cell

**3. Resource Scheduling:**

Among all other schedulers, Proportional Fair (PF) is one scheduling algorithm designed for serving optimum throughput in correspondence with a balanced user fairness index (FI) while allocating resources among multiple users. According to user priority, PF can provide a maximum data rate with little trade-off with fairness. Diving total bandwidth $B$ into sub-bands $S$, PF algorithm schedules resources based on the following equation[19]:

$$T_k(t+1) = \left(1 - \frac{1}{t_c}\right)T_k(t) + \frac{1}{t_c}\sum_{s=1}^{S} R_{s,k}(t) \quad (2)$$

where $T_k(t)$ is the average throughput, $R_{s,k}(t)$ is the data rate of user of subband $s$. The parameter $t_c$ is used to achieve sound trade-off between user throughput and the fairness factor.

In the case of Round Robin (RR) scheduler, it uses a simpler approach of allocating resources among the users in a cyclic manner ensuring the best fairness index without taking into account the channel quality. Resources are allocated based on a priority function $P$ with user data rate $R$ and throughput $T$[12]:

$$P = \frac{T^{\alpha}}{R^{\beta}} \quad (3)$$

Where $\alpha$ and $\beta$ are parameters to adjust corresponding fairness index. For RR, the parameters $\alpha = 0$ and $\beta = 1$.

## 4. Key Performance Indicators:

Average UE Throughput: Under a wireless coverage being operated by MU-MIMO, users are generally scattered all over the region with dynamic mobility. Under a certain BS coverage, distance of all the UEs from that certain BS is non-identical to each other, resulting a diverse range of user specific SINR and throughput. Thus, measuring the average UE throughput can express much about the present channel quality of the network, determined by[12]:

$$T_{avg} = \frac{\sum_{k=1}^{n} T_k}{n} \quad (4)$$

Where the total throughput is defined by $T_k$ for the $k^{th}$ user while the total number of users is defined by n.

Cell Edge Throughput: For a certain UE stationed at the edges of a macro cell, the huge distance along with inter-cell-interference (ICI) degrade the corresponding SINR by dropping down to the lowest. In this scenario, providing a minimum data rate necessary to avoid call drop occurrence during handovers. The cell edge throughput is measured from the $5^{th}$ percentile of the UE throughput ECDF.

Spectral Efficiency: How the data is being transmitted over a certain bandwidth with due efficacy can be delineated by the parameter, spectral efficiency and it can be expressed by[12]:

$$S = \frac{\sum_{k=1}^{n} T_k}{W} \quad (5)$$

Fairness Index: Fairness Index determines the distribution procedure of available resources among different users. Using Jain's fairness index described in [20], for $n$ users, corresponding fairness index can be determined by:

$$J(T) = \frac{[\sum_{k=1}^{n} T_k]^2}{n[\sum_{k=1}^{n} T_k^2]} \quad (6)$$

Where, the average throughput of $k^{th}$ user is defined by $T_k$.

## III. SIMULATION

Using LTE system level simulator, under CLSM MIMO scheme, different performance parameters are evaluated using various antenna combination: 2×2, 2×3, 2×4, 4×2, 4×3 and 4×4. Corresponding simulations are carried out in two phases with two different scheduling algorithms: PF and RR. Implementing user mobility effect using random walk model, overall, 570 UEs are scattered all over the geometric area in the simulation with 10 UEs in each sector. The random walk model implies that a variable, UE, independently moves an arbitrary step away from its previous value in each period. To ensure better performance with high MCS level, Mutual Information Based Effective SINR Mapping (MIESM) has been applied during simulation[21]. 20 TTIs and 50 TTIs have been considered for the simulations of RR and PF, respectively. Equation (9) represents an urban environment with macroscopic pathloss model for the corresponding macro cells[22]:

$$L = 40(1 - 40 * 10^{-3} h_{BS})log_{10}(R) - 18log_{10}(h_{BS})$$
$$+21log_{10}(f) + 80 dB \quad (7)$$

Where the separation between a BS and UE is defined by $R$ in kilometers, carrier frequency is defined by $f_c$ in MHz and height of an antenna is represented by $h_{BS}$ in meters. Additional parameters are charted in Table-II.

TABLE II – SIMULATION PARAMETERS FOR NETWORK MODEL

| Simulation Parameters | |
|---|---|
| Channel Model | WINNER+ |
| Frequency | 2.45 GHz |
| Bandwidth | 20 MHz |
| Antenna Combination | 2×2,2×3,2×4,4×2,4×3,4×4 |
| Transmission Mode | CLSM |
| BS Height | 20 m |
| BS Power | 45 db |
| Receiver Height | 1.5 m |
| Antenna Azimuth Offset | 30° |
| Antenna Gain | 15 dBi |
| BS Transmitter Power | 45 dBm |
| User Velocity | 0 – 120 kmph |
| Simulation Time | 50 TTI |

## IV. RESULTS AND DISCUSSION

As illustrated in Fig.2(a-b), a decline in average UE throughput occurs with uniform velocity increase. In Fig.2(a), under PF, for varied user mobility, the antenna combinations containing higher number of receiver antenna than the corresponding transmit antenna i.e., the set of 2×2,2×3,2×4 performs better than the set of 4×2,4×3,4×4. As the PF schedules resources based on current channel quality, the poor SINR experienced at high velocity causes the degradation of average throughput close to zero. With RR schedular, as depicted in Fig.2(b), at low velocity with better SINR, the 4×4 provides the best service. But the velocity increases, the 2×4 combination outperforms other diversities, due to the fact that the more receiver antennas are active than transmit antennas, the higher the probability of successful reception for receiver UE at poor channel quality.

In the case of cell edge throughput, depicted in Fig.3(a-b), combinations with higher number of transmit antennas compared to the number of receiver antenna provide decent throughput initially but after a while, drastic decline occurs as

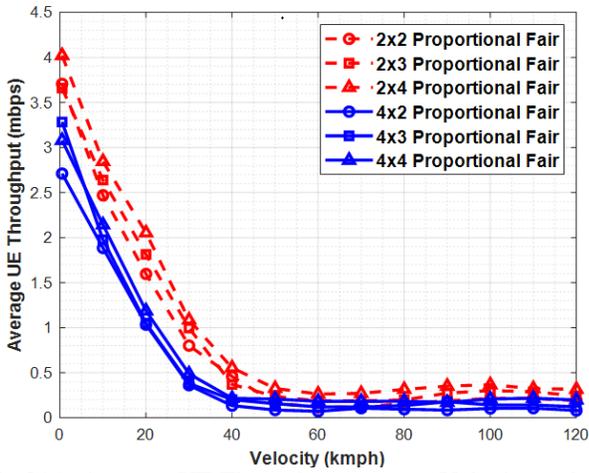
Fig.2(a): Average UE Throughput vs. user Velocity under PF scheduling algorithm

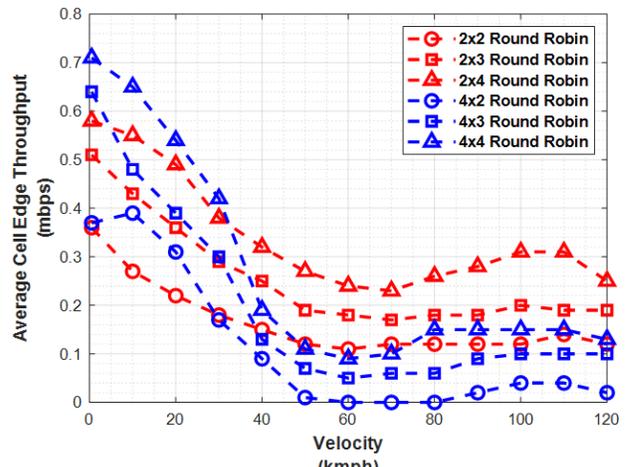
Fig.3(b): Average Cell Edge Throughput vs. user Velocity under RR scheduling algorithm

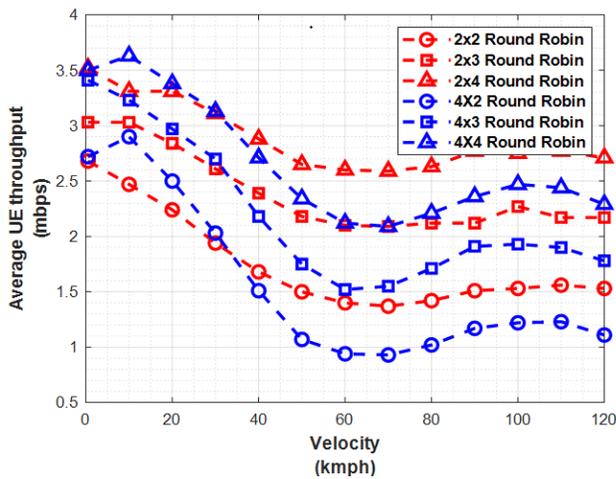
Fig.2(b): Average UE Throughput vs. user Velocity under RR scheduling algorithm

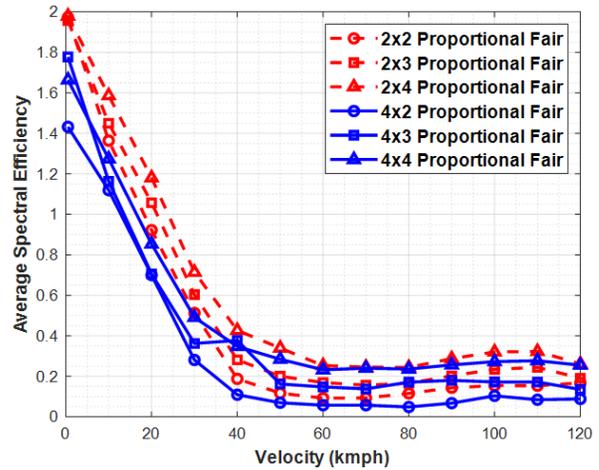
Fig.4(a): Average Spectral Efficiency vs. user Velocity under PF scheduling algorithm

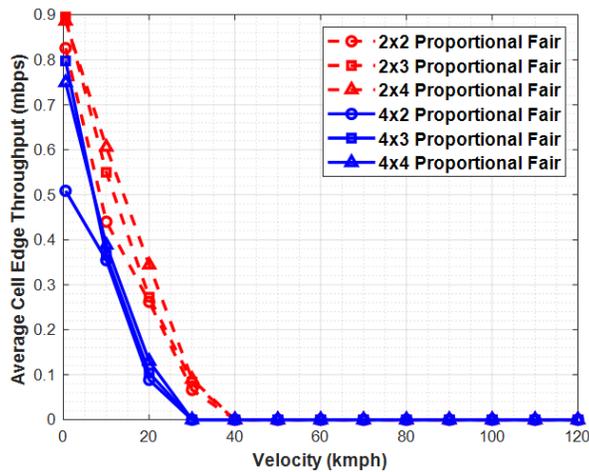
Fig.3(a): Average Cell Edge Throughput vs. user Velocity under PF scheduling algorithm

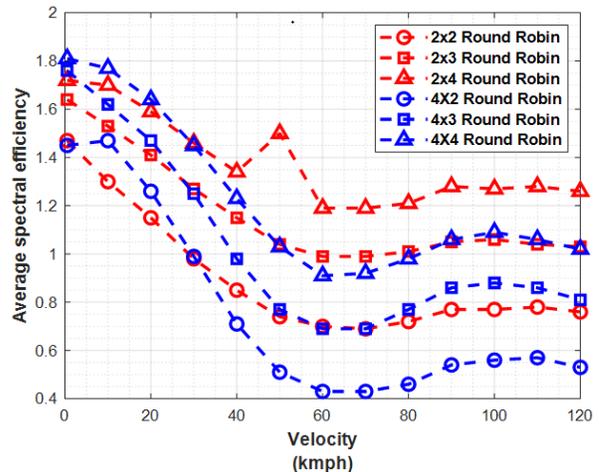
Fig.4(b): Average Spectral Efficiency vs. user Velocity under RR scheduling algorithm

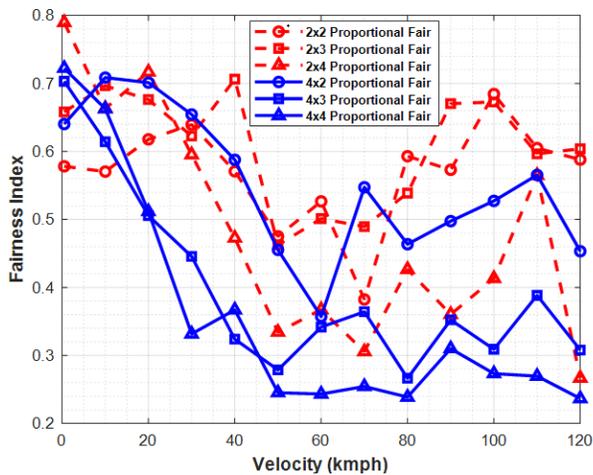

Fig.5(a): Fairness Index vs. user Velocity under PF scheduling algorithm

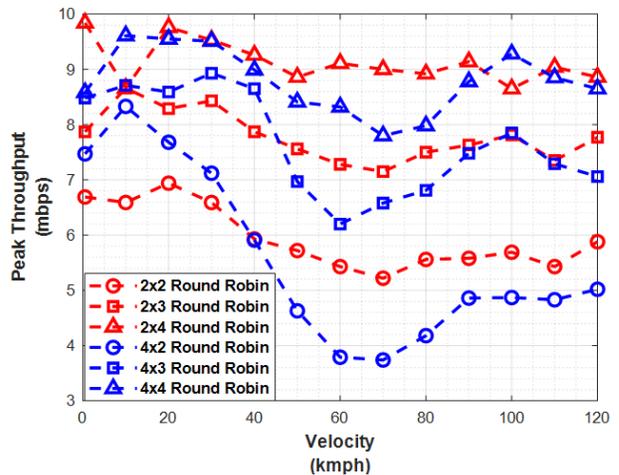

Fig.6(b): Peak Throughput vs. user Velocity under RR scheduling algorithm

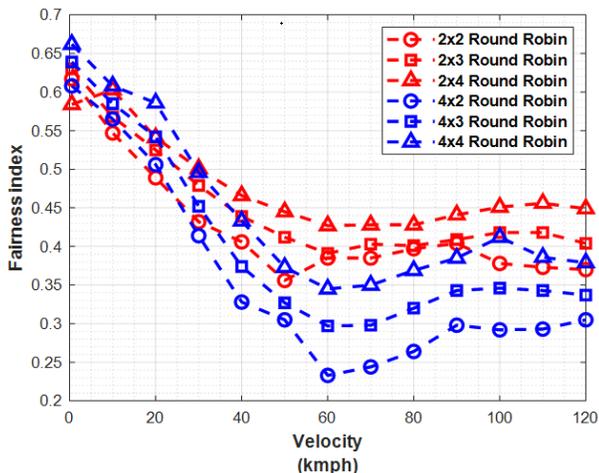

Fig.5(b): Fairness Index vs. user Velocity under RR scheduling algorithm

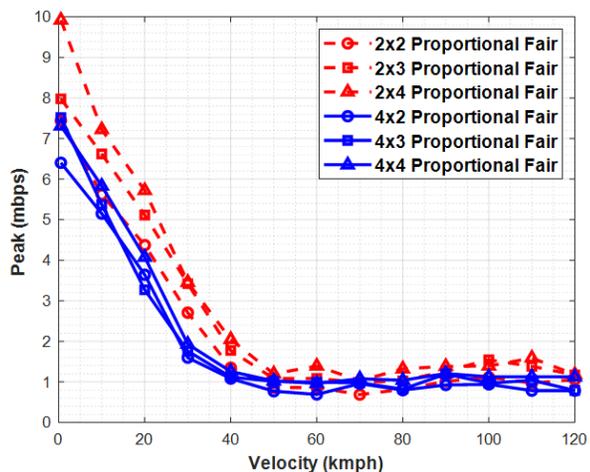

Fig.6(a): Peak Throughput vs. user Velocity under PF scheduling algorithm

SINR worsens at high velocity. In Fig.3(a), PF schedular provides decent throughput at low velocity while the set of the 2×2, 2×3 and 2×4 provides better service than others. After 40kmph, drastic decline occurs in the ongoing throughput, due to the poor SINR at high velocity. In Fig.3(b), with RR, the cell edge throughput drops with velocity increment but ultimately, it provides sufficient throughput to conduct successful handover, with that, the 2×4 combination triumphs over all.

With velocity increase, sharp drop in spectral efficiency is observed in Fig.4(a), operated with PF schedular and among other diversities, the 2×4 performs best in every occasion. With RR, in Fig.4(b), the 4×4 performs well at low velocity, but later, 2×4 and 2×3 come out with the best service than other diversities as they provide greater spectral efficiency for a given bandwidth with higher throughput at high velocity.

Observations in Fig.5(a-b) show deterioration in fairness index for various antenna combinations working under different scheduling algorithms as the velocity increases gradually. For both PF and RR, combinations with higher reception antennas i.e., 2×2, 2×3 and 2×4, with overall higher throughput quality, ensure better fairness index than other combinations especially at high UE mobility. Although, in comparison with Fig.5(a), RR generally delivers much better service in comparison with the PF, illustrated in Fig.5(b), due to cyclic manner of uniform resource allocation of RR among UEs without taking into account the channel condition and corresponding SINR.

Fig.6(a-b) indicate that, for both PF and RR, increasing the number of receiver antenna proportionately boosts up the peak throughput and corresponding network efficacy. In both results, achieving peak throughput using 2×4 combination above other diversities validate the fact, considering a wide range of user velocity scenario. It can also be observed that, as PF schedular distributes resources based on SINR, the peak gets quite low at high velocity due to the effect of poor channel condition at that range of velocity. On the contrary, as RR schedules resources in uniform manner, even at high velocity, quite similar amount

of peak throughput can be achieved in comparison with low velocity scenario, which is quite high.

## V. CONCLUSION

Downlink performance of a wireless network gets significantly affected by random and adverse change in the channel condition resulting from high velocity of cellular network users. In this paper, we propose to introduce antenna diversity in spatial multiplexing MIMO transmission scheme by operating greater number of reception antenna than the number of transmit antenna to improve the downlink performance of high velocity users. To conduct the study, a 4×4 MIMO system has been implemented for circumventing simulation complexity. Observations indicate the compatibility of this technique with any resource schedular and transmission scheme. After observing the performance of various transmit and receiver antenna combinations, the results show the fact that a steady data rate increment is achieved for a higher number of receiver antennas than the number of transmit antennas. This technique holds the potential for resolving real-life issues, like call drops and low data rate to be experienced by UEs traveling through high-speed transportation systems e.g. in Dhaka Metro MRT, and can be implemented in large scale antenna system for example in massive MIMO technology with minimal complexity.